\documentclass[twocolumn,dvipdfmx]{jpsj3}
\usepackage{graphicx}
\def\S{Sect. }
\def\deltac{\delta_{\rm c}}
\def\Ns{N_{\rm s}}
\def\Ms{M_{\rm s}}
\def\v#1{\mib #1}
\def\epsgtilde{{\tilde{\epsilon}_{\rm G}}}

\def\H{{\mathcal H}}
\newcommand{\bra}[1]{\left\langle {#1} \right\vert}
\newcommand{\bbra}[1]{\left.\left\langle {#1} \right\vert\right\vert}
\newcommand{\kket}[1]{\left\vert\left\vert {#1} \right.\right\rangle}
\newcommand{\ket}[1]{\left\vert {#1} \right\rangle}

\title
{
Ground-State Phases of Alternating-Bond $S = 1$ Diamond Chains
}

\author
{
Kazuo Hida\thanks{E-mail address: hida@mail.saitama-u.ac.jp}
}

\inst
{Professor Emeritus, Division of Material Science, Graduate School of Science and Engineering, \\ 
Saitama University, Saitama, Saitama, 338-8570\\ 
}

\recdate
{
}

\abst
{
The ground-state phases of alternating-bond spin-1 diamond chains are investigated. 
 Each ground state consists of an array of spin clusters separated by singlet dimers owing to an infinite number of local conservation laws. If no singlet dimers are present, the ground state is equivalent to that of a spin chain with infinite length.For strong frustration, we find a series of quantum phase transitions as in the case of alternating-bond mixed diamond chains with spins 1 and 1/2. For intermediate frustration, we find the nonmagnetic Haldane or dimer phases according to whether the bond alternation is weak or strong. For weak frustration and weak bond alternation, we find the ferrimagnetic states with spontaneous magnetizations $m=1/6$ and 1/3 per site. The ferrimagnetic state with $m=1/6$ is accompanied by a spontaneous translational symmetry breakdown. This phase vanishes for strong bond alternation.
}
\begin{document}
\maketitle
\section{Introduction}

The quantum effects in low-dimensional frustrated magnets have been extensively studied in recent condensed matter physics.\cite{intfrust,diep} Various exotic quantum phases emerge from the interplay of quantum fluctuation and frustration. From a theoretical point of view, it is remarkable that the ground states of a class of models are written down exactly owing to the frustration. A well-known example is the Majumdar-Ghosh model whose ground state is a dimer state with a spontaneous breakdown of translational symmetry.\cite{mg} The diamond chain\cite{Takano-K-S,tsh,ht2017,kiku2,kiku3} is another example of a model with an exact ground state. Since the diamond chain has an infinite number of local conservation laws, the ground states can be classified analytically. For strong frustration, each ground state consists of a periodic array of spin clusters separated by singlet dimers owing to these local conservation laws. With the decrease of frustration, the length of each spin cluster increases. For weak frustration, various nonmagnetic or ferrimagnetic states are realized depending on the magnitudes of the constituent spins and/or the exchange parameters. \cite{Takano-K-S,tsh,ht2017}

In the real materials with a diamond-chain structure, the lattice distortion is inevitable. In addition to the well-known natural mineral azurite,\cite{kiku2,kiku3} which is a spin-1/2 distorted diamond chain, it has been recently pointed out that 
 [Ni$_3$(OH)$_2$(O$_2$C-C$_2$H$_2$-CO$_2$)(H$_2$O)$_4$]$_2$H$_2$O\cite{guillou} can be regarded as a spin-1 distorted
 diamond chain.\cite{kuni,kiku1} Although the exchange parameters are fixed for real materials, even more exotic quantum states emerge in the whole parameter space of distorted diamond chains theoretically. 

The ground states of spin-1 diamond chains(S1DC) without distortion have been investigated in Refs. \citen{Takano-K-S} and \citen{ht2017}. In addition to the spin cluster ground states, the nonmagnetic Haldane state and the ferrimagnetic state with spontaneous magnetization $m=1/6$ and 1/3 are found. A series of quantum phase transitions take place between these ground-state phases with different periodicities and with or without a spontaneous translational symmetry breakdown. 
 
In the previous work\cite{hida2019}, we have investigated the effect of two types of lattice distortions that do not preserve the local conservation laws. In the present work, we consider the effect of the bond alternation depicted in Fig. \ref{lattice}. In contrast to the distortions discussed in Ref. \citen{hida2019}, the bond alternation preserves the same local conservation laws as the undistorted case.

This paper is organized as follows. 
 {In \S 2}, the model Hamiltonian is presented. 
 {In \S 3}, the ground-state phase diagram is determined analytically and numerically. 
The last section is devoted to summary and discussion.

\section{Hamiltonian}

\begin{figure}[t] 
\centerline{\includegraphics[width=6cm]{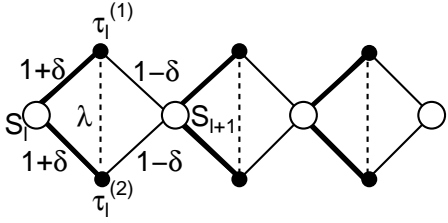}}
\caption{Structure of an alternating-bond S1DC. $S=\tau^{(1)}=\tau^{(2)}=1$.
The exchange parameters are denoted by $\lambda$, $1+\delta$, and $1-\delta$, 
where $\lambda$ and $\delta$ control the strengths of frustration and bond alternation, respectively.}
\label{lattice}
\end{figure}

 The alternating-bond S1DC is described by the Hamiltonian 
\begin{align}
{\mathcal H} = &\sum_{l=1}^{L} \left[(1+\delta)\v{S}_{l}(\v{\tau}^{(1)}_{l}+\v{\tau}^{(2)}_{l}) \right. 
\nonumber\\
&+(1-\delta)(\v{\tau}^{(1)}_{l}+\v{\tau}^{(2)}_{l})\v{S}_{l+1}
+ \left. \lambda\v{\tau}^{(1)}_{l}\v{\tau}^{(2)}_{l}\right] , 
\label{hama}
\end{align}
where $\v{S}_{l}$ and $\v{\tau}^{(\alpha)}_{l}\ (\alpha=1,2)$ are spin 1 operators. The number of the unit cells is denoted by $L$, and the total number of sites $\Ns$ is $3L$. 
 The Hamiltonian has three different exchange parameters, namely, 
$1+\delta$, $1-\delta$, and $\lambda$. Here, $\delta$ represents the strength of the bond alternation, and $\lambda$ controls the frustration as depicted in Fig. \ref{lattice}. 
In the case of $\delta = 0$, Eq.~(\ref{hama}) reduces to 
the Hamiltonian of S1DC without bond alternation.\cite{Takano-K-S,ht2017} 

The Hamiltonian (\ref{hama}) has a series of local conservation laws. 
To see it, we rewrite Eq. (\ref{hama}) 
 in the form, 
\begin{align}
\H &= \sum_{l=1}^{L} \H_l,\label{ham2}\\
\H_l&=(1+\delta)\v{S}_{l}\v{T}_{l}+(1-\delta)\v{T}_{l}\v{S}_{l+1}
+ \frac{\lambda}{2}\left(\v{T}^2_{l}-4\right), 
\end{align}
where the composite spin operators $\v{T}_l$ are defined as 
\begin{align}
\v{T}_{l} \equiv \v{\tau}^{(1)}_{l}+\v{\tau}^{(2)}_{l} 
\quad (l = 1, 2, \cdots ,L). 
\end{align}
Then, it is evident that 
\begin{align}
[\v{T}_l^2, {\mathcal H}] = 0 \quad (l = 1, 2, \cdots , L). 
\end{align}
Thus, we have $L$ conserved quantities $\v{T}_l^2$ for all $l$, even in the presence of bond alternation $\delta$. 
By defining the magnitude $T_l$ of the composite spin $\v{T}_l$ by $\v{T}_l^2 = T_l (T_l + 1)$, we have a 
 set of good quantum numbers $\{T_l; l=1,2,...,L\}$ where $T_l=$ 0,1, and 2. 
The total Hilbert space of the Hamiltonian (\ref{ham2}) consists of 
separated subspaces, each of which is specified by 
a definite set of $\{T_l\}$, i.e., a sequence of 0, 1, and 2. 
 A pair of spins with $T_l=0$ is a singlet dimer. 
 A cluster including $n$ successive $T_l\neq 0$ pairs bounded by two dimers is called a cluster-$n$. 

 Thus,  a cluster-$n$ is equivalent to the ground state of an alternating-bond antiferromagnetic Heisenberg chain consisting of $2n+1$ effective spins $\v{S}_l$ and $\v{T}_l$. 
 We follow the terminology of Ref. \citen{tsh} to call the ground state consisting of an array of cluster-$n$'s separated by dimers as dimer-cluster-$n$ (DC$n$) phase. The phases without dimers are called DC$\infty$ phases.

\section{Ground-State Phase Diagram}

\subsection{Ground states for $\lambda>2$}

\begin{figure} 
\centerline{\includegraphics[width=7cm]{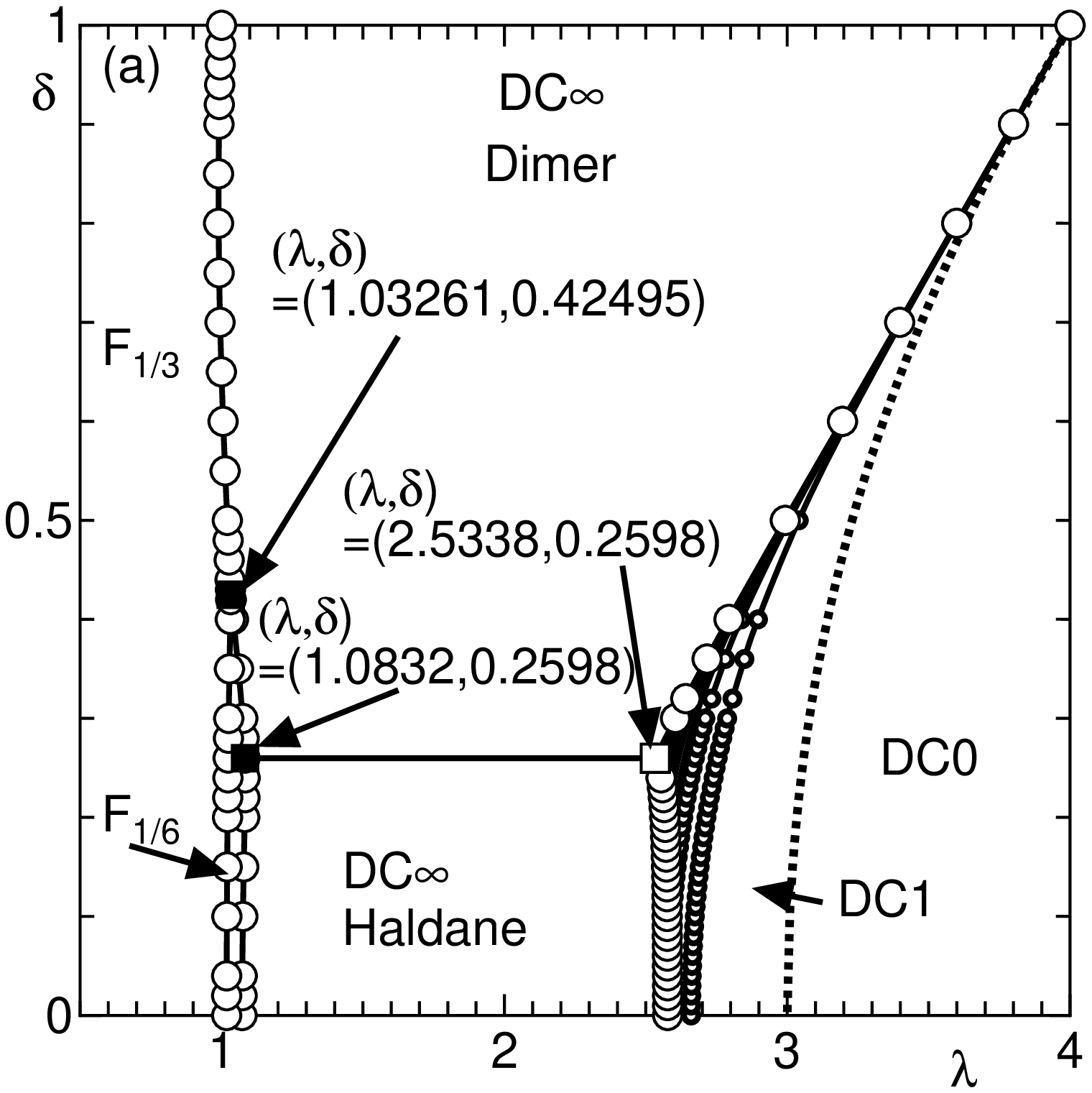}}
\centerline{\includegraphics[width=7cm]{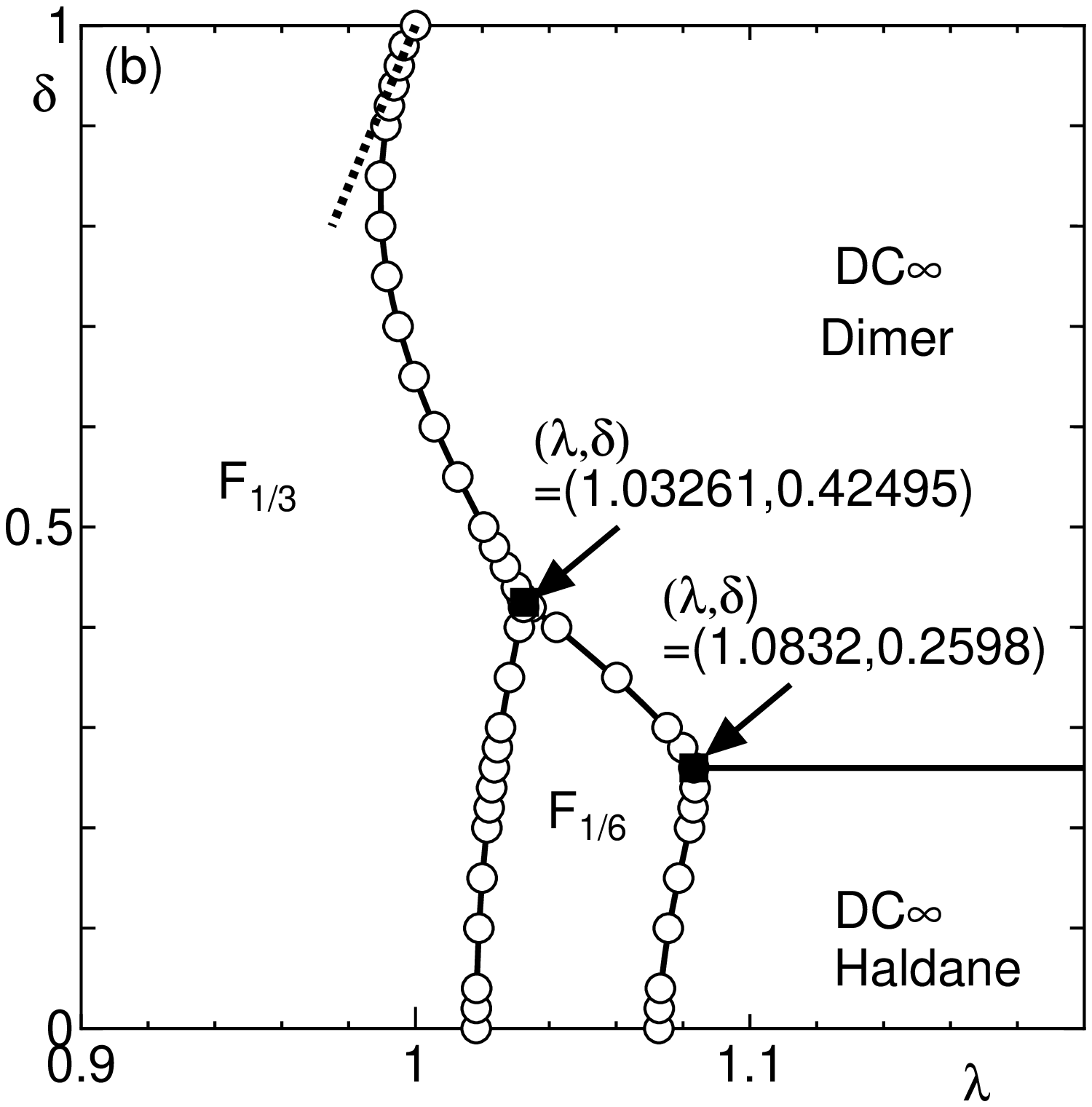}}
\caption{(a) Overall ground-state phase diagram. The region $\lambda >2$ is reproduced from Ref. \citen{hts_alt_mdc}. Small open circles are the phase boundaries between DC$n$ and DC$(n-1)$ phases. For $n \geq 4$, these phase boundaries are only shown by solid lines to avoid complications. Other phase boundaries are shown by the large open circles and solid lines. The DC0-DC1 phase boundary calculated exactly in Ref. \citen{hts_alt_mdc} is shown by the thick dotted line. The open square is the accumulation point of infinite series of critical points at $\delta=\deltac$.  The filled squares are triple points. (b) The enlarged phase diagram in the small $\lambda$ region. The thick dotted line is the approximate relation (\ref{eq:decouple}) for $ \delta \simeq  1$}
\label{phase}
\end{figure}

For $\lambda>2$, the state with $T_l=2$ is not allowed as in the case of undistorted S1DCs. This can be proven as follows:

Following Ref. \citen{Takano-K-S}, let us consider the ground state of the Hamiltonian of a single diamond $\H_l$ 
with $T_l=2$. According to the Lieb-Mattis theorem\cite{Lieb-Mattis}, the ground state is a singlet. By elementary manipulations, we can show that the following state $\ket{G}_l$ is the ground state of $\H_l$.
\begin{align}
\ket{G}_l&=\sqrt{\frac{1}{5}}\Big[\ket{-1}_{S_{l+1}}\ket{-1}_{S_l}\ket{2}_{T_{l}}
+\ket{1}_{S_{l+1}}\ket{1}_{S_l}\ket{-2}_{T_{l}}\nonumber\\
&+\sqrt{\frac{{2}}{{3}}}\ket{0}_{S_{l+1}}\ket{0}_{S_l}\ket{0}_{T_{l}}\nonumber\\
&-\frac{1}{\sqrt{2}}\left(\ket{-1}_{S_{l+1}}\ket{0}_{S_l}\ket{1}_{T_{l}}
+\ket{0}_{S_{l+1}}\ket{-1}_{S_l}\ket{1}_{T_{l}}\right.\nonumber\\
&\left.+\ket{0}_{S_{l+1}}\ket{1}_{S_l}\ket{-1}_{T_{l}}+\ket{1}_{S_{l+1}}\ket{0}_{S_l}\ket{-1}_{T_{l}}\right)
\nonumber\\
&+\frac{1}{\sqrt{6}}\ket{1}_{S_{l+1}}\ket{-1}_{S_l}\ket{0}_{T_{l}}+\frac{1}{\sqrt{6}}\ket{-1}_{S_{l+1}}\ket{1}_{S_l}\ket{0}_{T_{l}}\Big],
\end{align}
where $\ket{S_l^z}_{S_l}$ and $\ket{T_l^z}_{T_l}$ are the eigenstates of $S_l^z$ and $T_l^z$, respectively. In this state, we can show that 
\begin{align}
\v{S}_l\v{T}_l&=\v{S}_{l+1}\v{T}_l=-3.\label{eq:st}
\end{align}
Hence, we find the ground-state energy $E_l$ of $\H_l$ with $T_l=2$ as
\begin{align}
E_l(T_l=2)&=-6 +\lambda.
\end{align}
Obviously, $E_l(T_l=0)=-2\lambda$. Since these are independent of $\delta$, the conclusion that the state with $T_l=2$ is not allowed in the ground state for $\lambda >2$ remains valid even for $\delta \neq 0$ following the proof of Lemma 1 of Ref. \citen{Takano-K-S}. Hence, the ground-state phase diagram for $\lambda >2$ is identical to that of the mixed diamond chain with $(S,\tau)=(1,1/2)$ discussed in Ref. \citen{hts_alt_mdc} as shown in Fig. \ref{phase}(a). The main features are summarized as follows: 

The phase diagram consists of DC$n$ phases with finite $n$ and DC$\infty$ phases corresponding to the Haldane and dimer phases.  The DC$\infty$ phase is the Haldane phase or the dimer phase according to whether $\delta < \deltac\simeq 0.2598$ or $\delta > \deltac$, since the Hamiltonian (\ref{ham2}) reduces to the alternating-bond spin-1 Heisenberg chain.\cite{affleck1985,Affleck-Haldane1987,Kato-Tanaka1994,Yamamoto1994,Totsuka1995,Kitazawa-Nomura1997}  For $\delta=\deltac$ and $\delta \simeq 1$ the analytical results suggest the presence of an infinite series of phase transitions between DC$n$$(n=0, ...,\infty)$ phases with varying $\lambda$. Hence, we speculate that they should take place in the entire range of $\deltac \leq \delta \leq 1$. For $\delta=0$, only the DC$n$ phases for $n \leq 3$ and Haldane phase are realized.\cite{ht2017} For small $\lambda$, the ground state is the DC$\infty$ state that corresponds to the Haldane state for $\delta < \deltac$ and to the dimer state for $\delta > \deltac$. 
The physical quantities characterizing these phases are low temperature magnetic susceptibility and residual entropy.\cite{hts_alt_mdc} The behaviors of these quantities are also the same as those  described in Sect. 5.3 of Ref. \citen{hts_alt_mdc}.

\subsection{Ground-state phase diagram for $\lambda <2$}
\subsubsection{Possible candidates of ground states}
\label{sec_cand}
As in the case of $\delta=0$,\cite{ht2017} we employ the finite size DMRG method to calculate the ground-state energies of the Hamiltonian (\ref{ham2}) with open boundary condition  up to $L=13$ for all possible configurations $\{T_l\}$. For $\lambda <2$, we find that the possible configurations that give the lowest energies are (a) $\forall l, T_l=1$, (b) $\forall l, (T_{2l},T_{2l+1})=(1,2)$ or (2,1), (c) $\forall l, T_{l}=2$, and the configurations that differ from them only locally up to the system size $L=13$. From this observation, we choose the above three configurations (a)-(c) as most plausible candidates of the configurations of $T_l$'s in the DC$\infty$ ground states.
We will also examine some of the states with longer periodicities numerically. 
 For the case (a), the ground state is in the Haldane phase or the dimer phase according to whether $\delta < \deltac$ or $\delta > \deltac$\cite{affleck1985,Affleck-Haldane1987,Kato-Tanaka1994,Yamamoto1994,Totsuka1995,Kitazawa-Nomura1997}. 
 Thus, we expect the following four different phases in the DC$\infty$ phase.

\begin{enumerate}
\item Haldane phase (H phase) : 
$\{T_l\}=\{\dot{1}\}$ and $\delta < \deltac$.
\item Dimer phase (D phase) : 
$\{T_l\}=\{\dot{1}\}$ and $\delta > \deltac$.
\item Ferrimagnetic phase with $M=\Ms/6$ (F$_{1/6}$ phase) : 
$\{T_l\}= \{\dot{1}\dot{2}\}.$\label{f6}
\item Ferrimagnetic phase with $M=\Ms/3$ (F$_{1/3}$ phase) : 
$\{T_l\}=\{\dot{2}\}$.
\end{enumerate}
Here, $\{\dot{T_1}\cdots\dot{T}_l\}$ denotes the configuration consisting of a periodic array of the sequence $T_1\cdots T_l$ over the whole chain. The values of the spontaneous magnetizations in the ferrimagnetic phases can be identified from the Lieb-Mattis theorem\cite{Lieb-Mattis} .

\subsubsection{Approximation for $\delta \simeq 1$}

We decompose the Hamiltonian (\ref{ham2}) into two parts
\begin{align}
\H&=\H_0+\H_1,
\end{align}
where
\begin{align}
\H_0&=\sum_{l=1}^L \H_{0l},\ \ 
\H_1=\sum_{l=1}^L \H_{1l},\\
\H_{0l}&=(1+\delta)\v{S}_l\v{T}_l+\frac{\lambda}{2}[T_l(T_l+1)-4], \\ 
\H_{1l}&=(1-\delta)\v{S}_{l+1}\v{T}_l.
\end{align}
We denote the ground state of $\H_{0l}$ with total spin $S^{\rm tot}_l$, where $\v{S}^{\rm tot}_l\equiv \v{S}_l+\v{T}_l$, as $\kket{S^{\rm tot}_l,S^{{\rm tot} z}_l}_{T_lS_{l}}$, and corresponding eigenvalue as $E_{0,l}(S^{\rm tot}_l)$. Depending on the values of $T_l (=1, 2)$, these are given as follows:
\begin{enumerate}
\item $T_l=2$:

The ground state is a triplet state that has total spin $S^{\rm tot}_l=1$. They are expressed in terms of the states $\ket{{S}^z_l}_{S_l}$ and $\ket{{T}^z_l}_{T_{l}}$ as.
\begin{align}
\kket{1,1}_{T_lS_{l}}&=\sqrt{\frac{3}{5}}\left(\ket{-1}_{S_l}\ket{2}_{T_{l}}
-\frac{1}{\sqrt{2}}\ket{0}_{S_l}\ket{1}_{T_{l}}\right.\nonumber\\
&+\left.\frac{1}{\sqrt{6}}\ket{1}_{S_l}\ket{0}_{T_{l}}\right),\label{eq:kket11}\\
\kket{1,0}_{T_lS_{l}}&=\sqrt{\frac{3}{10}}\Big(\ket{-1}_{S_l}\ket{1}_{T_{l}}
-\frac{2}{\sqrt{3}}\ket{0}_{S_l}\ket{0}_{T_{l}}\nonumber\\
&+\ket{1}_{S_l}\ket{-1}_{T_{l}}\Big),\label{eq:kket10}\\
\kket{1,-1}_{T_lS_{l}}&=\sqrt{\frac{3}{5}}\left(\ket{1}_{S_l}\ket{-2}_{T_{l}}
-\frac{1}{\sqrt{2}}\ket{0}_{S_l}\ket{-1}_{T_{l}}\right.\nonumber\\
&+\left.\frac{1}{\sqrt{6}}\ket{-1}_{S_l}\ket{0}_{T_{l}}\right),\label{eq:kket1m1}
\end{align}
and the corresponding energy is given by $E_{0l}(1)=-3(1+\delta)+\lambda$.

\item $T_l=1$

The ground state is a singlet state $\kket{0,0}_{T_lS_{l}}$ with a total spin $S^{\rm tot}_l=0$ and the corresponding energy is given by $E_{0l}(0)=-2(1+\delta)-\lambda$. The wave function is not presented since it is unnecessary for the following calculations.
\end{enumerate}

For $\delta=1$, the candidates of the ground states found in section \ref{sec_cand} can be constructed from the above local states as follows: 

The $\{\dot{2}\}$ phase that corresponds to the ferrimagnetic ground state with magnetization $1/3$ per spin is given by 
\begin{align}
\ket{F_{1/3}}=\prod_{l=1}^{L}\kket{1,1}_{T_lS_{l}},
\end{align}
if the spontaneous magnetization is taken along the $z$-axis. The energy of this phase is given by
\begin{align}
E_{1/3,0}/L=-3(1+\delta)+\lambda.
\end{align}
The $\{\dot{1}\}$ phase that corresponds to the nonmagnetic dimer ground state is given by 
\begin{align}
\ket{D}=\prod_{l=1}^{L}\kket{0,0}_{T_lS_{l}}.
\end{align}
The energy of this phase is given by
\begin{align}
E_{D,0}/L=-2(1+\delta)-\lambda.
\end{align}

The $\{\dot{1}\dot{2}\}$ phase that corresponds to the ferrimagnetic ground state with magnetization $1/6$ per spin is given by 
\begin{align}
\ket{F_{1/6}}&=\prod_{l=1}^{L/2}\kket{0,0}_{T_{2l-1}S_{2l-1}}\kket{1,1}_{T_{2l}S_{2l}},
\end{align}
or
\begin{align}
\ket{F_{1/6}}&=\prod_{l=1}^{L/2}\kket{1,1}_{T_{2l-1}S_{2l-1}}\kket{0,0}_{T_{2l}S_{2l}},
\end{align}
if the spontaneous magnetization is taken along the $z$-axis. The energy of this phase is given by
\begin{align}
E_{1/6,0}/L=-\frac{5}{2}(1+\delta).
\end{align}

Next, we calculate the energy of each ground state up to the first order with respect to the perturbation $\H_1$.
Among the terms in $\H_{1l}$, ${T}^{\pm}_l{S}^{\mp}_{l+1}$ has no diagonal elements within the ground state subspace of $\H_0$. Hence, we only consider the term ${T}^{z}_l{S}^{z}_{l+1}$. 
By elementary manipulations, we find
\begin{align}
\bbra{1,1}_{T_lS_{l}}T^z_l\kket{1,1}_{T_lS_{l}}&
=\frac{3}{2},\\
\bbra{1,1}_{T_{l+1}S_{l+1}}S^z_{l+1}\kket{1,1}_{T_{l+1}S_{l+1}}&
=-\frac{1}{2}.
\end{align}
Hence, we have 
\begin{align}
E_{1/3,1}&=\bra{F_{1/3}}\H_1\ket{F_{1/3}}=-\frac{3L}{4}(1-\delta).
\end{align}
Since $\ket{D}$ is nonmagnetic, expectation values of $T^z_l$ and $S^z_{l+1}$ vanish. Hence, we have
\begin{align}
E_{{\rm D},1}&=\bra{D}\H_1\ket{D}=0.
\end{align}

In $\ket{F_{1/6}}$, one of the groups of spins ($\v{S}_l$,$\v{T}_l$) or ($\v{S}_{l+1}$,$\v{T}_{l+1}$) connected by $\H_{1l}$ is always nonmagnetic. Hence, the expectation value of $T^z_lS^z_{l+1}$ vanishes. Thus, we find
\begin{align}
E_{1/6,1}&=\bra{F_{1/6}}\H_1\ket{F_{1/6}}=0.
\end{align}
Then, the ground state energies of the whole diamond chain in these three phases are given by
\begin{align}
E_{1/3}/L&=-3(1+\delta)-\frac{3}{4}(1-\delta)+\lambda,\\
E_{1/6}/L&=-\frac{5}{2}(1+\delta),\\
E_{\rm D}/L&=-2(1+\delta)-\lambda,
\end{align}
up to the first order in $\H_1$. The state $\ket{F_{1/6}}$ always has higher energy than the states $\ket{F_{1/3}}$ and $\ket{D}$ within the present approximation. Hence, it is not realized for $\delta \simeq 1$. Thus, the phase transition between the phases $\ket{F_{1/3}}$ and $\ket{D}$ takes place at
\begin{align}
\lambda_{\rm DF_{1/3}}(\delta)=1-\frac{1}{8}(1-\delta). 
\label{eq:decouple}
\end{align}

\subsubsection{Numerical calculation}

Following the argument in Ref. \citen{ht2017}, for general values of $\delta$, the phase boundaries are determined from the ground-state energy per spin $\epsgtilde(\{T_l\},\delta)$ of an infinite-length Heisenberg chain with bond alternation $\delta$,
\begin{align}
\tilde{\H}=\sum_{l=1}^L \left[(1+\delta)\v{S}_l\v{T}_l+(1-\delta)\v{T}_l\v{S}_{l+1}\right],
\end{align}
with the corresponding configuration of $\{T_l\}$ in the following way:
\begin{enumerate}
\item Haldane-F$_{1/6}$ phase boundary for $\delta < \deltac$:
\begin{align}
 {\lambda}_{\rm HF_{1/6}}(\delta)&=2\epsgtilde(\{\dot{1}\},\delta) -2\epsgtilde(\{\dot{1}\dot{2}\},\delta).
\end{align}
\item Dimer-F$_{1/6}$ phase boundary for $\delta > \deltac$:
\begin{align}
 {\lambda}_{\rm DF_{1/6}}(\delta)
&=2\epsgtilde(\{\dot{1}\},\delta) -2\epsgtilde(\{\dot{1}\dot{2}\},\delta).
\end{align}
\item F$_{1/6}$-F$_{1/3}$ phase boundary:
\begin{align}
{\lambda}_{\rm F_{1/6}F_{1/3}}(\delta)
=2\epsgtilde(\{\dot{1}\dot{2}\},\delta)-2\epsgtilde(\{\dot{2}\},\delta).
\end{align}
\item D-F$_{1/3}$ direct transition phase boundary:
\begin{align}
 {\lambda}_{\rm DF_{1/3}}(\delta)
=\epsgtilde(\{\dot{1}\},\delta)-\epsgtilde(\{\dot{2}\},\delta).
\end{align}
\end{enumerate}
The energy $\epsgtilde(\{T_l\},\delta)$ can be calculated from the increment of the ground-state energy in each step of the iDMRG calculation.\cite{iDMRG,DMRG_rev} The phase diagram is shown in Fig. \ref{phase}. The number of states $\chi$ kept in each iDMRG step ranged from 240 to 480. Convergence with respect to $\chi$ is confirmed.

There are two triple points for $\lambda <2$. Among them, the D-F$_{1/3}$-F$_{1/6}$ triple point is $(\lambda,\delta)\simeq(1.03261,0.42495)$ and the D-H-F$_{1/6}$ triple point is  $(\lambda,\delta)=(\lambda_{\rm c},\deltac)$ where $\lambda_{\rm c}\simeq 1.0832$.

So far, we restricted ourselves to the candidate ground states with configurations (a)-(c). In addition to these most plausible candidates, we also examined the states with $\{\dot{1}...\dot{1}\dot{2}\}$ structures that have longer periodicities. We find these states are not ground states for almost all values of $\delta$ and $\lambda$. Nevertheless, in the close neighborhood of the point  $(\lambda_{\rm c},\deltac)$, we obtained numerical results suggesting the presence of the F$_{1/9}$ and  partial ferrimagnetic ground states\cite{pferri_s_h,furuya2014} in a tiny region of width $\Delta\lambda \sim 10^{-3}$. However, the complete analysis of this region requires extensive high precision numerical calculations for all possible configurations of $\{T_l\}$. Hence, we leave the full investigation of this region for future studies.

\section{Summary and Discussion}

 The ground-state phases of the alternating-bond S1DC are investigated. Owing to the local conservation laws, the ground states are rigorously constructed, once the ground states of corresponding Heisenberg chains are known. For $\lambda >2$, the local state $T_l=2$ is not allowed. Hence, the ground state is identical to that of the (1,1/2) alternating-bond mixed diamond chain discussed in Ref. \citen{hts_alt_mdc}. Each ground-state phase is described as a DC$n$ phase consisting of a uniform array of cluster-$n$'s that are equivalent to the ground states of spin-1 alternating-bond Heisenberg chains of length $2n+1$. The nonmagnetic DC$\infty$ phases that correspond to Haldane or dimer phases also appear for intermediate $\lambda$.

 For small $\lambda$, the ground states are the DC$\infty$ states. Especially for $\lambda <2$, the local state with $T_l=2$ is allowed. Hence, ferrimagnetic phases with $m=1/6$ and 1/3 are also realized in addition to the nonmagnetic Haldane or dimer phases. The ferrimagnetic phase with $m=1/6$ is accompanied by the spontaneous translational symmetry breakdown.

For $\delta \simeq 1$, we have carried out the perturbation analysis with respect to $1-\delta$. In this regime, only the dimer phase and ferrimagnetic phase with $m=1/3$ are realized and no spontaneous translational symmetry breakdown takes place. 

In the close neighborhood of the point  $(\lambda_{\rm c},\deltac)$, we obtained numerical results suggesting the presence of a narrow partial ferrimagnetic phase. This might be related to the presence of the adjacent Haldane-dimer critical line with a Luttinger-liquid ground state along the line $\delta=\deltac$ and $\lambda > \lambda_{\rm c}$, since the partial ferrimagnetic phase can be regarded as a spontaneously magnetized Luttinger liquid\cite{furuya2014}.

Thus, we found that the introduction of the bond alternation to the S1DC gives rise to a rich variety of quantum phases and phase transitions. The experimental realization of the S1DC has been reported in Refs. \citen{kuni} and \citen{kiku1}, although the distortion pattern is different. We hope the rich phase diagram predicted in the present paper stimulates the synthesis of other S1DC materials in the near future. 

\acknowledgments

We thank K. Takano for stimulating comments and discussion. The numerical diagonalization program is based on the package TITPACK ver.2 coded by H. Nishimori. A part of the numerical computation in this work has been carried out using the facilities of the Supercomputer Center, Institute for Solid State Physics, University of Tokyo and Yukawa Institute Computer Facility at Kyoto University.


\begin{thebibliography}{50}

\bibitem{intfrust} {\it Introduction to Frustrated Magnetism: Materials, Experiments, Theory}, ed. C. Lacroix, P. Mendels, and F. Mila 
 (Springer Series in Solid-State Sciences, Springer, Heidelberg, 2011).
\bibitem{diep} {\it Frustrated Spin Systems}, ed. H. T. Diep, (World Scientific, Singapore, 2013) 2nd ed. 
\bibitem{mg} C. K. Majumdar and D. K. Ghosh, J. Math. Phys. {\bf 10}, 1399 (1969). 
\bibitem{Takano-K-S}
K. Takano, K. Kubo, and H. Sakamoto, J. Phys.: Condens. Matter {\bf 8}, (1996) 6405. 
\bibitem{tsh}
K. Takano, H. Suzuki, and K. Hida, Phys. Rev. B {\bf 80}, 104410 (2009). 
\bibitem{ht2017} K. Hida and K. Takano, J. Phys. Soc. Jpn. {\bf 86}, 033707 (2017).
\bibitem{kiku2}H. Kikuchi, Y. Fujii, M. Chiba, S. Mitsudo, T. Idehara, T. Tonegawa, K. Okamoto, T. Sakai, T. Kuwai, and H. Ohta, Phys. Rev. Lett. {\bf 94}, 227201 (2005).
\bibitem{kiku3} H. Kikuchi, Y. Fujii, M. Chiba, S. Mitsudo, T. Idehara, T. Tonegawa, K. Okamoto, T. Sakai, T. Kuwai T, K. Kindo, A. Matsuo, W. Higemoto, K. Nishiyama, M. Horovi\'c, and C. Bertheir, Prog. Theor. Phys. Suppl. {\bf 159}, 1 (2005).
\bibitem{guillou} N. Guillou, S. Pastre, C. Livage, and G. F\'erey, Chem. Commun. 2002, 2358.
\bibitem{kuni} K. Kunieda, Master Thesis, University of Fukui (2016)[in Japanese].
\bibitem{kiku1} H. Kikuchi, private communication.
\bibitem{hida2019} K. Hida, J. Phys. Soc. Jpn. {\bf 88}, 074705 (2019).
\bibitem{Lieb-Mattis} E. Lieb and D. Mattis, J. Math. Phys. {\bf 3}, 749 (1962).
\bibitem{hts_alt_mdc} K. Hida, K. Takano, and H. Suzuki, J. Phys. Soc. Jpn. {\bf 79}, 044702 (2010).
\bibitem{affleck1985} I. Affleck, Nucl. Phys. B{\bf 257} [FS 14], 397 (1985). 
\bibitem{Affleck-Haldane1987} I. Affleck and F. D. M. Haldane, Phys. Rev. B{\bf 36}, 5291 (1987).
\bibitem{Kato-Tanaka1994} Y. Kato and A. Tanaka, J. Phys. Soc. Jpn. {\bf 63}, 1277 (1994).
\bibitem{Yamamoto1994} S. Yamamoto, J. Phys. Soc. Jpn. {\bf 63}, 4327 (1994); Phys. Rev. B{\bf 51}, 16128 (1995). 
\bibitem{Totsuka1995} K. Totsuka, Y. Nishiyama, N. Hatano, and M. Suzuki, J. Phys.: Condens. Matter {\bf 7}, 4895 (1995).
\bibitem{Kitazawa-Nomura1997} A. Kitazawa and K. Nomura, J. Phys. Soc. Jpn. {\bf 66}, 3944 (1997).
\bibitem{iDMRG} I. P. McCulloch, arXiv:0804.2509 (2008).
\bibitem{DMRG_rev} U. Schollw\"ock, Ann. Phys. {\bf 326}, 96 (2011).
\bibitem{pferri_s_h} K. Sekiguchi and K. Hida, J. Phys. Soc. Jpn. {\bf 86}, 084706 (2017) and references therein.
\bibitem{furuya2014} S. C. Furuya and T. Giamarchi, Phys. Rev. B{\bf 89}, 205131 (2014).
\end{thebibliography}
\end{document}